\documentclass[aps,prl,preprint,nopacs,superscriptaddress]{revtex4}
\usepackage{chngcntr}
\usepackage{graphicx}
\usepackage{verbatim}
\usepackage{mathrsfs}
\usepackage{gensymb}
\usepackage{color}
\usepackage{ulem}

\usepackage{amsmath,amsfonts,amssymb}
\usepackage{graphicx}

\newcommand{\bee}{\begin{equation}}
\newcommand{\ee}{\end{equation}}
\def\3{2.8in}    
\def\2{2.5in}
\def\4{3.0in}\def \beq {\begin{equation}}
\def \eeq {\end{equation}}
\begin{document}

\title{Type-II Topological Dirac Semimetals: Theory and Materials Prediction (VAl$_3$ family)}
\author{Tay-Rong~Chang$^*$}
\affiliation {Department of Physics, National Tsing Hua University, Hsinchu 30013, Taiwan}
\author{Su-Yang Xu\footnote{These authors contributed equally to this work.}$^{\dag}$}\affiliation {Laboratory for Topological Quantum Matter and Spectroscopy (B7), Department of Physics, Princeton University, Princeton, New Jersey 08544, USA}
\author{Daniel S. Sanchez}\affiliation {Laboratory for Topological Quantum Matter and Spectroscopy (B7), Department of Physics, Princeton University, Princeton, New Jersey 08544, USA}

\author{Shin-Ming Huang}
\affiliation{Department of Physics, National Sun Yat-Sen University, Kaohsiung 804, Taiwan}

\author{Guoqing Chang}\affiliation{Centre for Advanced 2D Materials and Graphene Research Centre National University of Singapore, 6 Science Drive 2, Singapore 117546}\affiliation{Department of Physics, National University of Singapore, 2 Science Drive 3, Singapore 117542}
\author{Chuang-Han Hsu}\affiliation{Centre for Advanced 2D Materials and Graphene Research Centre National University of Singapore, 6 Science Drive 2, Singapore 117546}
\author{Guang Bian}\affiliation {Laboratory for Topological Quantum Matter and Spectroscopy (B7), Department of Physics, Princeton University, Princeton, New Jersey 08544, USA}
\author{Ilya Belopolski}\affiliation {Laboratory for Topological Quantum Matter and Spectroscopy (B7), Department of Physics, Princeton University, Princeton, New Jersey 08544, USA}

\author{Zhi-Ming Yu}\affiliation{School of Physics, Beijing Institute of Technology, Beijing 100081, China} \affiliation {Research Laboratory for Quantum Materials, Singapore University of Technology and Design, Singapore 487372, Singapore}
\author{Xicheng Xu}
\affiliation{International Center for Quantum Materials, School of Physics, Peking University, China}\affiliation{Collaborative Innovation Center of Quantum Matter, Beijing,100871, China}
\author{Cheng Xiang}
\affiliation{International Center for Quantum Materials, School of Physics, Peking University, China}\affiliation{Collaborative Innovation Center of Quantum Matter, Beijing,100871, China}
\author{Shengyuan A. Yang}\affiliation{Research Laboratory for Quantum Materials, Singapore University of Technology and Design, Singapore 487372, Singapore}
\author{Titus Neupert}\affiliation {Department of Physics, University of Zurich, 190, CH-8052, Switzerland, Winterthurerstrass}
\author{Horng-Tay Jeng}
\affiliation{Department of Physics, National Tsing Hua University, Hsinchu 30013, Taiwan}
\affiliation{Institute of Physics, Academia Sinica, Taipei 11529, Taiwan}
\author{Hsin Lin$^{\dag}$}
\affiliation{Centre for Advanced 2D Materials and Graphene Research Centre National University of Singapore, 6 Science Drive 2, Singapore 117546}
\affiliation{Department of Physics, National University of Singapore, 2 Science Drive 3, Singapore 117542}
\author{M. Zahid Hasan\footnote{Corresponding authors (emails): suyangxu@princeton.edu, nilnish@gmail.com, mzhasan@princeton.edu }}\affiliation {Laboratory for Topological Quantum Matter and Spectroscopy (B7), Department of Physics, Princeton University, Princeton, New Jersey 08544, USA}

\pacs{}

\date{\today}

\begin{abstract}

The discoveries of Dirac and Weyl semimetal states in spin-orbit compounds led to the realizations of elementary particle analogs in table-top experiments. In this paper, we propose the concept of a three-dimensional type-II Dirac fermion and identify a new topological semimetal state in the large family of transition-metal icosagenides, MA$_3$ (M=V, Nb, Ta; A=Al, Ga, In). We show that the VAl$_3$ family features a pair of strongly Lorentz-violating type-II Dirac nodes and that each Dirac node consists of four type-II Weyl nodes with chiral charge $\pm{1}$ via symmetry breaking. Furthermore, we predict the Landau level spectrum arising from the type-II Dirac fermions in VAl$_3$ that is distinct from that of known Dirac semimetals. We also show a topological phase transition from a type-II Dirac semimetal to a quadratic Weyl semimetal or a topological crystalline insulator via crystalline distortions. The new type-II Dirac fermions, their novel magneto-transport response, the topological tunability and the large number of compounds make VAl$_3$ an exciting platform to explore the wide-ranging topological phenomena associated with Lorentz-violating Dirac fermions in electrical and optical transport, spectroscopic and device-based experiments.
\end{abstract}

\maketitle

The correspondence between condensed matter and high-energy physics has been a source of inspiration throughout the history of physics. Advancements in topological band theory have uncovered a new and profound relations that has enabled the realization of elementary relativistic fermions in crystals with unique topologically non-trivial properties \cite{Herring,Weyl,Ashvin_book, TI_book_2015, Na3Bi,Na3Bi_2,Na3Bi_3,Cd3As2_2,Cd3As2_4, Nagaosa, Wan2011,Burkov2011,TaAs_the,TaAs_the2,Hasan_TaAs,MIT_Weyl,TaAs_Ding,TS-1,TS-3,TS-4}. Specifically, the low-energy quasiparticle excitations of type-I Dirac semimetals \cite{Na3Bi,Na3Bi_2,Na3Bi_3,Cd3As2_2,Cd3As2_4, Nagaosa}, type-I Weyl semimetals \cite{Wan2011,Burkov2011,TaAs_the,TaAs_the2,Hasan_TaAs,MIT_Weyl,TaAs_Ding}, and topological superconductors \cite{TS-1,TS-3,TS-4} are the direct representations of relativistic Dirac, Weyl, and Majorana fermions, respectively. From an application perspective, what makes this realized connection with high-energy physics of importance and interest is the resulting broad range of topologically protected phenomena that can be potentially used for low-power electronics, spintronics, and robust qubits \cite{Nonlocal,Chiral_anomaly_ChenGF,Chiral_anomaly_Jia}. For these reasons, the type-I Dirac semimetal state in Na$_3$Bi \cite{Na3Bi_2,Na3Bi_3} and Cd$_3$As$_2$ \cite{Cd3As2_2,Cd3As2_4}, the type-I Weyl semimetal state in the TaAs family of crystals \cite{Hasan_TaAs,TaAs_Ding}, and the various topological superconductor candidates \cite{TS-1,TS-3,TS-4} have attracted tremendous interest. Very recently, a new line of thinking that looks for new topological quasiparticles beyond direct analogs in high-energy physics, has gained attention. Such an idea offers inroads into new topological phenomena that are not limited by the stringent constraints in high-energy physics \cite{Grushin,Bergholtz_1,Bergholtz_2}. A particularly interesting proposal is the prediction of type-II emergent Weyl fermions \cite{WT-Weyl}. Type-I Weyl fermions, which have been realized in the TaAs family of crystals, are the direct analogs of the massless relativistic Weyl fermion from high-energy. They respect Lorentz symmetry and have the typical conical dispersion. In contrast, type-II Weyl fermions are dramatically Lorentz symmetry breaking, which is manifest in a tilted-over cone in energy-momentum space \cite{WT-Weyl}. These Lorentz violating Weyl fermions can give rise to many new properties, such as a direction-dependent chiral anomaly \cite{FS_anomaly}, an anti-chiral effect of the chiral Landau level \cite{SAYang}, novel quantum oscillations due to momentum space Klein tunneling \cite{Klein_tun}, and a modified anomalous Hall conductivity \cite{QAH}. The novel type-II Weyl semimetal state has been recently predicted/confirmed in a number of 3D crystals \cite{WT-Weyl, MT-Weyl, WMTe2, MT-Weyl-2, WT-ARPES-0-0, WT-ARPES-0-1, WT-ARPES-0-2, LGA, TaS, TIT}.

Since the type-II behavior only relies on the fact that Lorentz invariance is not a necessary symmetry requirement in condensed matter physics, in solid-state crystals, Lorentz symmetry breaking is not limited to the type-II Weyl fermion and, in principle, can emerge in other particles, including the Weyl fermion's most closely related particle, the Dirac fermion. However, to date, three-dimensional (3D) type-II Dirac fermions remain entirely lacking. Here, for the first time, we propose the concept of the 3D type-II Dirac semimetal state and identify it in a large family of transition-metal icosagenides, MA$_3$ (M=V, Nb, Ta; A=Al, Ga, In).

{\bf Type-II Dirac fermions}

The VAl$_3$ family of compounds crystalizes in a body-centered tetragonal Bravais lattice with lattice constants $a = 3.78$ $\textrm{\AA}$ and $c = 8.322$ $\textrm{\AA}$ \cite{VAl3_str} and the space group $I4/mmm$ (No. $139$), as shown in Fig.~\ref{structure}\textbf{a}. In this structure, each Al atom is surrounded by four V atoms in two different local structures: a planer square and a tetrahedron geometry (Fig.~\ref{structure}\textbf{b}). Figure~\ref{structure}\textbf{c} shows the bulk Brillouin zone of the VAl$_3$ crystal.

Interestingly, without spin-orbit coupling (SOC), the conduction and valence bands cross to form a triply-degenerate point along the $\Gamma-Z$ line, which, upon the inclusion of SOC, evolves into the type-II Dirac node. We present a physical picture for the formation of the type-II Dirac fermions in VAl$_3$ (Figs. \ref{structure}\textbf{d-f}). The important symmetries are time-reversal ($\mathcal{T}$), space-inversion ($\mathcal{I}$), the $C_{4z}$ rotation, and the $\mathcal{M}_{x}$ mirror reflection. Since the $\Gamma-Z$ line lies in the $\mathcal{M}_{x}$ mirror plane and coincides with the $C_{4z}$ rotational axis, the electron states along $\Gamma-Z$ are required to be eigenstates of both operators. In the absence of SOC, the $C_{4z}$ operator has the following four eigenvalues, $\pm1$ and  $\pm{i}$, and we denote the corresponding eigenstates as $\tilde\psi_1$, $\tilde\psi_{-1}$, $\tilde\psi_i$, and $\tilde\psi_{-i}$, respectively. Under the $\mathcal{M}_{x}$ mirror reflection, $\tilde\psi_1$, $\tilde\psi_{-1}$ remain unchanged ($\mathcal{M}_x\tilde\psi_1=\tilde\psi_1$, $\mathcal{M}_x\tilde\psi_{-1}=\tilde\psi_{-1}$), whereas $\tilde\psi_i$ and $\tilde\psi_{-i}$ will transform into each other $\mathcal{M}_x\tilde\psi_i=\tilde\psi_{-i}$. Thus, as $\mathcal{M}_{x}$ and $C_{4z}$ do not commute in this subspace, $\psi_i$ and $\psi_{-i}$ must remain degenerate everywhere along the $\Gamma-Z$ line. In the case of VAl$_3$, our calculation shows that the observed triply-degenerate point is formed by the crossing between a $\tilde\psi_{-1}$ (singly-degenerate) band and a $\tilde\psi_{\pm{i}}$  (doubly-degenerate) band (Fig. \ref{structure}\textbf{d}). This triple degeneracy is protected because the presence of $C_{4z}$ rotational symmetry prohibits hybridization between states of different $C_{4z}$ eigenvalues. Now we explain how the triply-degenerate point evolves into the type-II Dirac node upon the inclusion of SOC. With SOC, the four eigenvalues of $C_{4z}$ become $e^{\frac{i\pi}{4}}$, $e^{3i\frac{\pi}{4}}$, $e^{5i\frac{\pi}{4}}$, and $e^{i\frac{7\pi}{4}}$, and we denote the corresponding eigenstates as $\psi_{e^{i\frac{\pi}{4}}}$, $\psi_{e^{i\frac{3\pi}{4}}}$, $\psi_{e^{i\frac{5\pi}{4}}}$, and $\psi_{e^{i\frac{7\pi}{4}}}$. Adding SOC also doubles the number of bands at play (Fig. \ref{structure}\textbf{e}). Along $\Gamma-Z$, the doubling of the bands can be tracked by their $C_{4z}$ eigenvalues. Specifically, we have $\tilde\psi_1\Rightarrow\psi_{e^{i\frac{7\pi}{4}}},\psi_{e^{i\frac{\pi}{4}}};\quad\tilde\psi_i\Rightarrow\psi_{e^{i\frac{\pi}{4}}},\psi_{e^{i\frac{3\pi}{4}}};\quad\tilde\psi_{-1}\Rightarrow\psi_{e^{i\frac{3\pi}{4}}},\psi_{e^{i\frac{5\pi}{4}}};\quad\tilde\psi_{-i}\Rightarrow\psi_{e^{i\frac{5\pi}{4}}},\psi_{e^{i\frac{7\pi}{4}}}$. Under the $\mathcal{M}_{x}$ mirror reflection, $\psi_{e^{i\frac{\pi}{4}}}$ and $\psi_{e^{i\frac{7\pi}{4}}}$ transform into each other ($\mathcal{M}_x\psi_{e^{i\frac{\pi}{4}}}=i\psi_{e^{i\frac{7\pi}{4}}}$), and $\psi_{e^{i\frac{3\pi}{4}}}$ and $\psi_{e^{i\frac{5\pi}{4}}}$ transform into each other ($\mathcal{M}_x\psi_{e^{i\frac{5\pi}{4}}}=i\psi_{e^{i\frac{3\pi}{4}}}$). Therefore, $\psi_{e^{i\frac{\pi}{4}}}$ and
$\psi_{e^{i\frac{7\pi}{4}}}$ are degenerate everywhere along $\Gamma-Z$ and the same is true for $\psi_{e^{i\frac{3\pi}{4}}}$ and $\psi_{e^{i\frac{5\pi}{4}}}$. Taking the above conclusions from symmetry analyses, We can understand the evolution from triply-degenerate node to the type-II Dirac node. Specifically, the $\tilde\psi_{-1}$ (singly-degenerate) band without SOC becomes a $\psi_{e^{i\frac{3\pi}{4}}}, \psi_{e^{i\frac{5\pi}{4}}}$ (doubly-degenerate) band with SOC. The $\tilde\psi_{-\pm{i}}$ (doubly-degenerate) band without SOC becomes two doubly-degenerate bands ($\psi_{e^{i\frac{\pi}{4}}},\psi_{e^{i\frac{7\pi}{4}}}$ and $\psi_{e^{i\frac{3\pi}{4}}}, \psi_{e^{i\frac{5\pi}{4}}}$) with SOC. Crossings between bands with the same $C_{4z}$ eigenvalues open a gap. On the other hand, crossings between bands with the different $C_{4z}$ eigenvalues remains gapless. This gives rise to the Dirac node of VAl$_3$ (Fig. \ref{structure}\textbf{f}).

We now present the calculated band structure of VAl$_3$ to reveal the Dirac node and its type-II character. Figure~\ref{band}\textbf{a} shows the calculated bulk band structure along high symmetry directions. We mark the energy gap between the lowest conduction and valence bands by the green shaded areas. The conduction and valence bands cross each other near the Fermi level at discrete $k$ point, revealing the semimetallic ground state. In Fig.~\ref{band}\textbf{d}, we show the zoomed-in view of the band structure near the crossing point. We see that, along the $\Gamma-Z$ direction, the conduction and valence bands cross each other, forming the Dirac node near the $Z$ point. Figures~\ref{band}\textbf{b,c} show the energy dispersion away from the Dirac node along all three $k$ directions. While the two bands have Fermi velocities of opposite signs along $k_x$ and $k_y$ directions, they have  velocities of the same sign along $k_z$. Moreover, the constant energy contour at the energy of the Dirac node consists of an electron pocket and a hole pocket touching at the Dirac node. These observations demonstrate the first type-II Dirac fermion semimetal state. We note that a few recent papers \cite{Isobe, Car, Schnyder} have discussed type-II linear band crossings in 2D. Here our focus is 3D. In fact, Weyl fermions are not allowed in 2D so linear band crossings in 2D such as graphene and the surface states of topological insulators are usually called ``Dirac'' but that mainly refers to the linear dispersions. Only in 3D, the distinction between Dirac and Weyl becomes well-defined, and carries a topological meaning because Weyl fermions have a nonzero chiral charge whereas 3D Dirac fermions do not.

\vspace{1.5cm}

{\bf Topological invariant and Fermi arc surface states.}

In order to understand the topological properties of the type-II Dirac semimetal state in VAl$_3$, we calculate the 2D $\mathbb{Z}_2$ invariant $\nu_{2\textrm{D}}$ and the mirror Chern number $n_{\mathcal{M}}$. A 2D $\mathbb{Z}_2$ invariant can be defined on any 2D $k$ plane that is time-reversal invariant and whose band structure has a full energy gap. In VAl$_3$, the 2D planes that satisfy these conditions include $k_{z}=0$ and $k_{z}=\pi$. Since the system preserves space-inversion symmetry, $\nu_{2\textrm{D}}$ can be computed through a parity analysis. In Table 1, we show the parity total eigenvalue of the occupied states for the eight time-reversal invariant momenta, from which we conclude that both the $k_{z}=0$ and $k_{z}=\pi$ planes have a trivial $\mathbb{Z}_2$ number ($\nu_{k_{z}=0}=\nu_{k_{z}=\pi} = 0$). A mirror Chern number $n_{\mathcal{M}}$ can be defined on any 2D $k$ plane that is invariant under a mirror reflection and whose band structure has a full energy gap. In VAl$_3$, only the $k_{z}=0$ plane satisfies these conditions. We computed the $n_{\mathcal{M}}$ associated with the $\mathcal{M}_{z}$ mirror plane by the Kubo formula, and our calculations show $n_{\mathcal{M}}=2$ for the $k_{z}=0$ plane. We note that existing Dirac semimetals Na$_3$Bi and Cd$_3$As$_2$ are known to possess a 2D $\mathbb{Z}_2$ ($\nu_{2D} = 1$) and a mirror Chern number ($n_{\mathcal{M}} = 1$), respectively \cite{Nagaosa}. Therefore, the result $n_{\mathcal{M}}=2$ indicates that the Dirac semimetal state in VAl$_3$ is topologically distinct from that of in Cd$_3$As$_2$ or Na$_3$Bi \cite{Nagaosa}.

\begin{table}
\caption{The product of parity eigenvalue of the occupied states for eight time-reversal symmetry momenta in BZ.}
 \label{table1}
 \begin{center}
  \begin{tabular}{cccc}
   \hline \hline
   $\Gamma$ & 2$X$ & $Z$ & 4$N$ \\
   \hline
   +        & +    & +   & -    \\
   \hline \hline
  \end{tabular}
 \end{center}
\end{table}

We now explore the existence of protected surface states in VAl$_3$ and their connection to the type-II Dirac nodes. In order to do so, we calculate the surface electronic structure of the (100) surface, where the two Dirac nodes are projected onto different $k$ locations in the surface BZ. Figures~\ref{ss}\textbf{d,e} show how the bulk BZ is projected onto the (100) surface. Due to the body-centered structural property,  VAl$_3$'s (100) surface BZ center $\bar{\Gamma}$ corresponds to the projection of both the $\Sigma-\Gamma$ line and the $\Sigma_{1}-Z$ line. Because the Dirac nodes are near the $Z$ point in the bulk, their surface projections are close to the $\bar{\Gamma}$ point. Figure~\ref{ss}\textbf{a} shows the energy dispersion of surface band structure along the $\bar{\Gamma}-\bar{N}$ ($k_z$) direction. The shaded areas represent the projected bulk bands whereas the distinct lines are the surface states. We observe the type-II bulk Dirac fermion cone in the form of a tilted-over cone near the $\bar{\Gamma}$ point ($k_z\simeq0.15\frac{2\pi}{c}$). We find surface states that emerge out of the Dirac node, suggesting the existence of Fermi arcs. In Figs.~\ref{ss}\textbf{b,c}, we present the surface constant energy contour at the energy of the bulk Dirac nodes. We see two pairs of Fermi arcs terminated onto a Dirac node. They start from the Dirac node and quickly merge onto the projected electron-like pocket. According to the current theoretical understanding \cite{Nagaosa}, the surface states in a Dirac semimetal are not required to connect the Dirac node because a Dirac node does not carry nonzero chiral charge. However, band structure calculations have now found Fermi arcs connecting Dirac nodes in all known Dirac semimetals including Na$_3$Bi, Cd$_3$As$_2$ \cite{Na3Bi,Na3Bi_2}, and VAl$_3$ (this work). Additionally, we notice that at $E\sim35$ meV above E$_F$, the normal surface states (Fig.~\ref{ss}\textbf{a}) are observed to accidentally cross with the Fermi arcs and form a surface (2D) type-II Dirac cone. Although such crossings are not guaranteed by any topological invariant, they are protected because the surface bands that cross have opposite $\mathcal{M}_y$ eigenvalues.

\vspace{1.5cm}

{\bf Topological phase transitions in VAl$_3$.}

To further showcase the novel physics that may be studied in VAl$_3$, we will now turn our attention towards investigating its topological phase transitions. Generically, a Dirac semimetal can be regarded as a critical point of different phases. By starting from the Dirac semimetal phase, it is possible to realize different topological phases by simply tuning various parameters. In Fig.~\ref{FS}\textbf{e} we show a cartoon illustration of the (001) surface Fermi surface of VAl$_3$. Interestingly, the mirror Chern number $n_{\mathcal{M}} = 2$ defined on the $k_{z}=0$ plane and the number of Fermi arcs found at each Dirac node are consistent with each other. We first show the topological phase transition from the type-II Dirac semimetal state to a topological crystalline insulator state. We break the $C_{4z}$ rotational symmetry by compressing the lattice along the $\hat{x}$ direction which makes $a\neq{b}$. This opens up a gap at VAl$_3$'s type-II Dirac nodes. Because of $n_{\mathcal{M}} = 2$ at $k_z=0$, the resulting insulating phase is a topological crystalline insulator with two Dirac surface states as shown in Fig.~\ref{FS}\textbf{d}. We now show two consecutive topological phase transitions which transform the type-II Dirac fermions first to quadratic double Weyl fermions with chiral charge $\pm2$ then to linear single Weyl fermions with chiral charge $\pm1$ . As shown in Figs.~\ref{FS}\textbf{e,f}, we apply an Zeeman field along the $k_z$ direction. This field breaks time-reversal symmetry but preserves the $C_4$ rotational symmetry. As a result, each type-II Dirac node is found to split into a pair of quadratic Weyl nodes with chiral charge of $\pm2$. Because of the $\pm2$ chiral charge, each quadratic Weyl node is required to have two Fermi arcs. Interestingly, the four Fermi arcs associated with each Dirac node naturally provide the Fermi arcs needed for the pair of quadratic Weyl nodes. Since the $C_4$ rotational symmetry is preserved, the quadratic Weyl nodes are still along the $C_4$ rotational ($k_z$) axis. The $k$ space distribution of the quadratic Weyl nodes (Fig.~\ref{FS}\textbf{e}) breaks time-reversal symmetry. As a result, as shown in Fig.~\ref{FS}\textbf{e}, any $(k_x, k_y)$ slice whose $k_z$ is between the immediate pair of quadratic Weyl nodes has a Chern number of $2$. The fact that the BZ carries nonzero chiral number suggests the existence of anomalous Hall current $\sigma_{xy}$ that arises from the quadratic Weyl nodes \cite{QAH_Weyl}. Figure~\ref{FS}\textbf{i} shows the calculated band structure along $k_z$ in the presence of the Zeeman field, where we see that the type-II Dirac fermion (Fig.~\ref{FS}\textbf{h}) indeed splits into a pair of type-II quadratic Weyl nodes. Each quadratic Weyl fermion disperses quadratically along $k_x$ and $k_y$ directions (Fig.~\ref{FS}\textbf{j}) but linearly along $k_z$ direction (Fig.~\ref{FS}\textbf{i}). In Figs.~\ref{FS}\textbf{f,g}, we further break the $C_4$ rotational symmetry. We find that the $C_4$ breaking splits each double Weyl nodes with chiral charge $\pm2$ into two single Weyl node with chiral charge $\pm1$. Therefore, a net number of four Weyl nodes are generated from a singe type-II Dirac node in VAl$_3$.  Depending on whether one compresses the lattice along $\hat{x}$ or $\hat{y}$ direction, the splitting of the double Weyl nodes will be along $k_x$ or $k_y$ direction. Figure~\ref{FS}\textbf{k-i} show the situation where the lattice was compressed along the $\hat{x}$ direction. We see that each quadric Weyl node split into two single Weyl nodes along $k_y$ direction.

\vspace{1.5cm}

{\bf Landau level spectrum}

We study the Landau level spectrum of the type-II Dirac fermions in VAl$_3$ (Fig. \ref{LL}). In the presence of an external magnetic field, electrons in 2D move in cyclotron orbits with quantized energy values, the Landau levels. In a 3D material, the Landau levels further gain dispersions along the direction of the magnetic field. The Landau level spectrum plays a key role in dictating the magneto-transport properties of materials. In Fig. \ref{LL}\textbf{a}, we show the Landau level spectrum for the type-II Dirac fermions in VAl$_3$. We see that, except the red lines which are the lowest Landau levels, all higher Landau levels are segregated into two groups (the conduction and valence bands) separated by an energy gap. The lowest Landau levels are found to connect the conduction and valence bands across the band gap. The existence of zeroth ($0^{\textrm{th}}$) Landau levels connecting the band gap is a signature of Dirac/Weyl fermions. In the case of a Weyl fermion, the $0^{\textrm{th}}$ Landau level is one chiral band (Fig.~\ref{LL}\textbf{c}). In the case of a Dirac fermion, the $0^{\textrm{th}}$ Landau levels are a pair of counter-propagating chiral bands as a Dirac fermion consists of a pair of Weyl fermions of opposite chirality. This can be easily seen in a type-I Dirac fermion shown in Fig.~\ref{LL}\textbf{a}. For the type-II Dirac fermion in VAl$_3$ (Fig.~\ref{LL}\textbf{b}), the two chiral bands have Fermi velocities of the same sign, which seems to contradict the above picture. This is because the band gap, i.e., the white region in Fig.~\ref{LL}\textbf{b} separating the conduction and valence bands, is heavily tilted due to the type-II character. If one tilts the band gap back to being horizontal, then the two $0^{\textrm{th}}$ Landau levels are counter-propagating.

We find that the type-II character in VAl$_3$ leads to a distinct response in its Landau level spectrum, i.e., the existence of a critical angle of the magnetic field, along which all Landau levels ``collapse'' into the same energy, giving rise to a large density of states \cite{Klein_tun, SAYang}. We start from the condition where the magnetic field is parallel to the tilting direction of the type-II Dirac cone, which is the $k_z$ direction in VAl$_3$. In this case, the electrons' cyclotron motions are within the $(k_x, k_y)$ plane. In the semiclassical picture, the electrons will trace the Fermi contour within this plane. Fig.~\ref{LL}\textbf{d} shows the band dispersion along $kk_x$. Since the dispersion has a typical conical shape, the constant energy contour is a closed loop independent of the chemical potential position (the top panel of Fig.~\ref{LL}\textbf{g}). We now vary the magnetic field direction within the $(k_x, k_z)$ plane. Figure\ref{LL}\textbf{e} shows the energy dispersion along $k_{\theta1}$ that is perpendicular to the magnetic field $B_{\theta1}$ (Fig.~\ref{LL}\textbf{g}). We see that the dispersion becomes a tilted cone, while the constant energy contour is a still closed loop (the middle panel of Fig.~\ref{LL}\textbf{g}). As we continue to tilt the magnetic field, there exists a critical angle, at which one of the bands becomes flat. This means that, to the lowest order ($k^1$ term in the $k\cdot{p}$ theory), the Fermi contour becomes non-closed (the bottom panel of Fig.~\ref{LL}\textbf{g}). Interestingly, our calculations show that, to the lowest order, all Landau levels collapse to the same energy. When considering higher order terms, the Landau levels are still confined to a very narrow energy window at this critical angle, leading to a large density of states (Fig.~\ref{LL}\textbf{f}). Based on the calculated band structure of VAl$_3$, we obtain a critical angle $0.247\pi$ (between the magnetic field and $k_z$) for the type-II Dirac fermions in VAl$_3$.

Finally, we consider the effects of exchange/Zeeman coupling of the magnetic field. Figures~\ref{LL}\textbf{h-k} show the Landau level spectrum with different magnitudes of the Zeeman field. It can be seen that, the inclusion of an Zeeman coupling moves the two crossings of the two counter-propagating chiral bands closer (Figs.~\ref{LL}\textbf{h,i}). This corresponds to process shown in Figs.~\ref{LL}\textbf{l-o}, where an Zeeman field in VAl$_3$ splits each type-II Dirac node into a pair of quadratic double Weyl nodes with opposite chiral charges $\pm2$ and increasing the Zeeman field will increase the splitting, which moves two of the four double Weyl nodes closer to the $\bar{\Gamma}$ point.  If one keeps increasing the Zeeman coupling, there exists a critical value of the Zeeman coupling, at which the two crossings meet at $k_z=0$ and at the same time the band gap closes (Fig.~\ref{LL}\textbf{j}). This corresponds to process shown in Fig. \ref{LL}\textbf{n}, where the two double Weyl nodes meet and annihilate at the $\bar{\Gamma}$ point. If one further increases the Zeeman coupling beyond this critical value, the band gap reopens and two co-propagating chiral bands appear (Fig. \ref{LL}\textbf{k}). This corresponds to scenario shown in Fig. \ref{LL}\textbf{o}. Only one pair of double Weyl nodes are left following the annihilation at $\bar{\Gamma}$. Since each double Weyl node has a chiral charge of $\pm2$, it contributes two co-propagating chiral Landau bands (Fig. \ref{LL}\textbf{k}).

The rich and novel Landau level spectra of the type-II Dirac fermions in VAl$_3$ uncovered above suggest distinct and novel magneto-transport phenomena that can be measured in electrical transport, optical transport, and scanning tunneling spectroscopic experiments.

\vspace{1.5cm}

\newpage

\clearpage
\begin{figure}
\centering
\includegraphics[width=16cm]{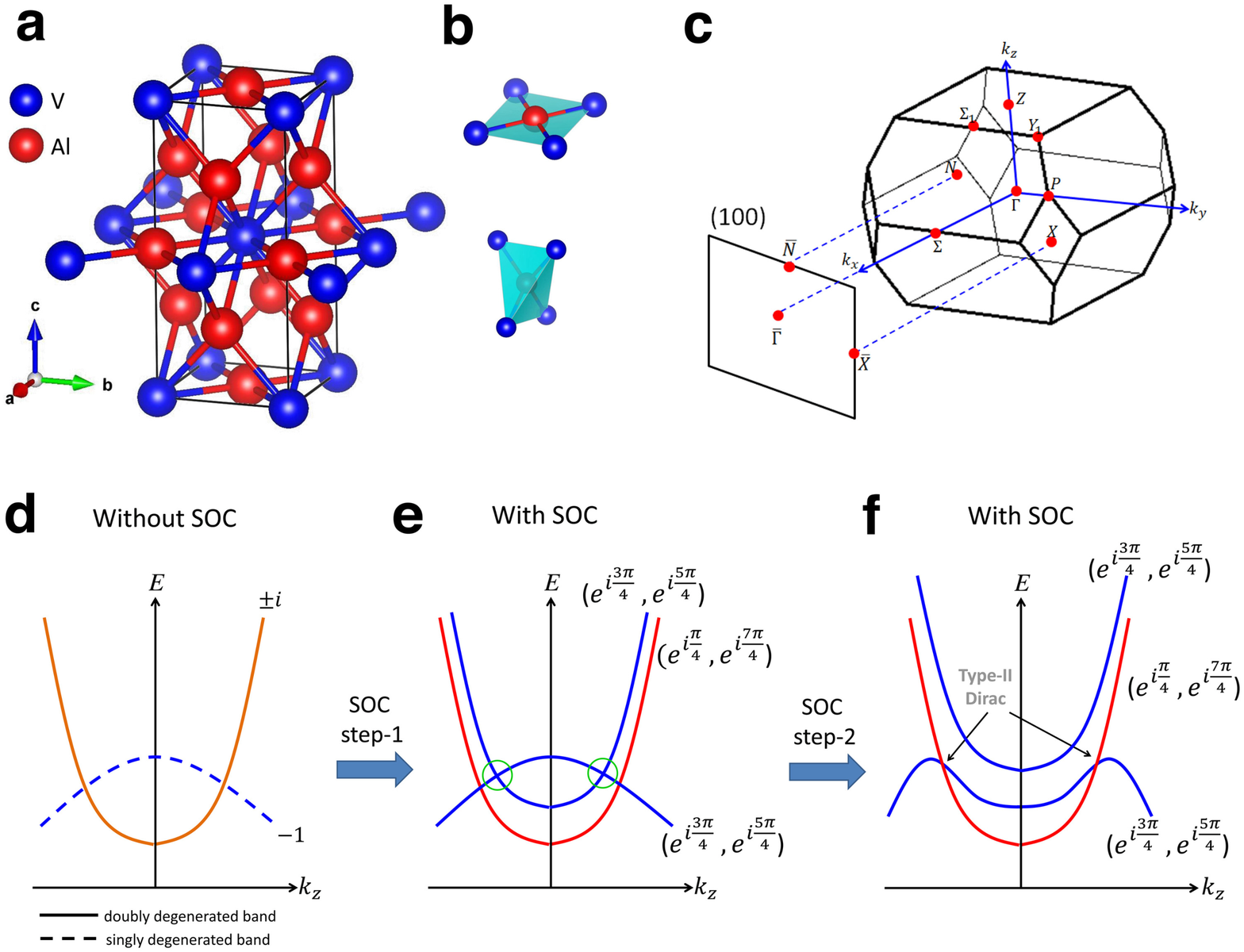}
\caption{
{\bf Symmetry origins for the type-II Dirac fermions in VAl$_3$.}
{\bf a,b,} VAl$_3$ crystalizes in a body centered tetragonal Bravais lattice with lattice constants $a = 3.78$ $\textrm{\AA}$  and $c = 8.322$ $\textrm{\AA}$ and space group $I4/mmm$ (No. $139$). The crystals unit cell is shown with the blue and red spheres representing the V and Al atoms, respectively. As can be seen, the crystal lattice has spatial inversion symmetry, $C_{4z}$ symmetry, and mirror reflection planes $\mathcal{M}_x$ and $\mathcal{M}_z$. {\bf c,} The bulk Brillouin zone (BZ) of VAl$_3$ with the relevant high-symmetry points marked with red dots. The (100) surface BZ is also shown with its projected high-symmetry points. {\bf d,} In the absence of SOC, the triply-degenerate point along the $\Gamma$-$Z$ direction is formed by a doubly degenerate conduction band (red band) and singly degenerate valence band (blue-dashed band). Defined by $C_{4z}$ operator, the $\pm{i}$ eigenvalues of the doubly-degenerate band and $-1$ eigenvalue of the singly degenerate bands is shown. {\bf e,} In the presence of SOC, every band becomes doubled. The initial doubly degenerate conduction band in the no SOC case gets turned into two doubly degenerate bands conduction bands (blue and red bands).The resulting new eigenvalues for the blue and red doubly degenerate conduction bands are ($e^{i\frac{3\pi}{4}}$, $e^{i\frac{5\pi}{4}}$) and ($e^{\frac{i\pi}{4}}$, $e^{i\frac{7\pi}{4}}$), respectively. }
\label{structure}
\end{figure}
\addtocounter{figure}{-1}
\begin{figure*}[t!]
\caption{The initial singly degenerate valence band in the no SOC case get turned into a doubly degenerate band (blue band) with eigenvalues ($e^{i\frac{3\pi}{4}}$, $e^{i\frac{5\pi}{4}}$). {\bf f,} Enclosed by green circles in panel (e) are the touching points between two doubly-degenerate blue bands with the same $C_{4z}$ eigenvalues, which results in these states becoming gapped. However, the two doubly degenerate red and blue bands do not share the same eigenvalues and, therefore, generate crossing points that are not gapped. The new generated crossing points are the type-II Dirac nodes.}
\label{structure}
\end{figure*}

\clearpage
\begin{figure}
\centering
\includegraphics[width=16cm]{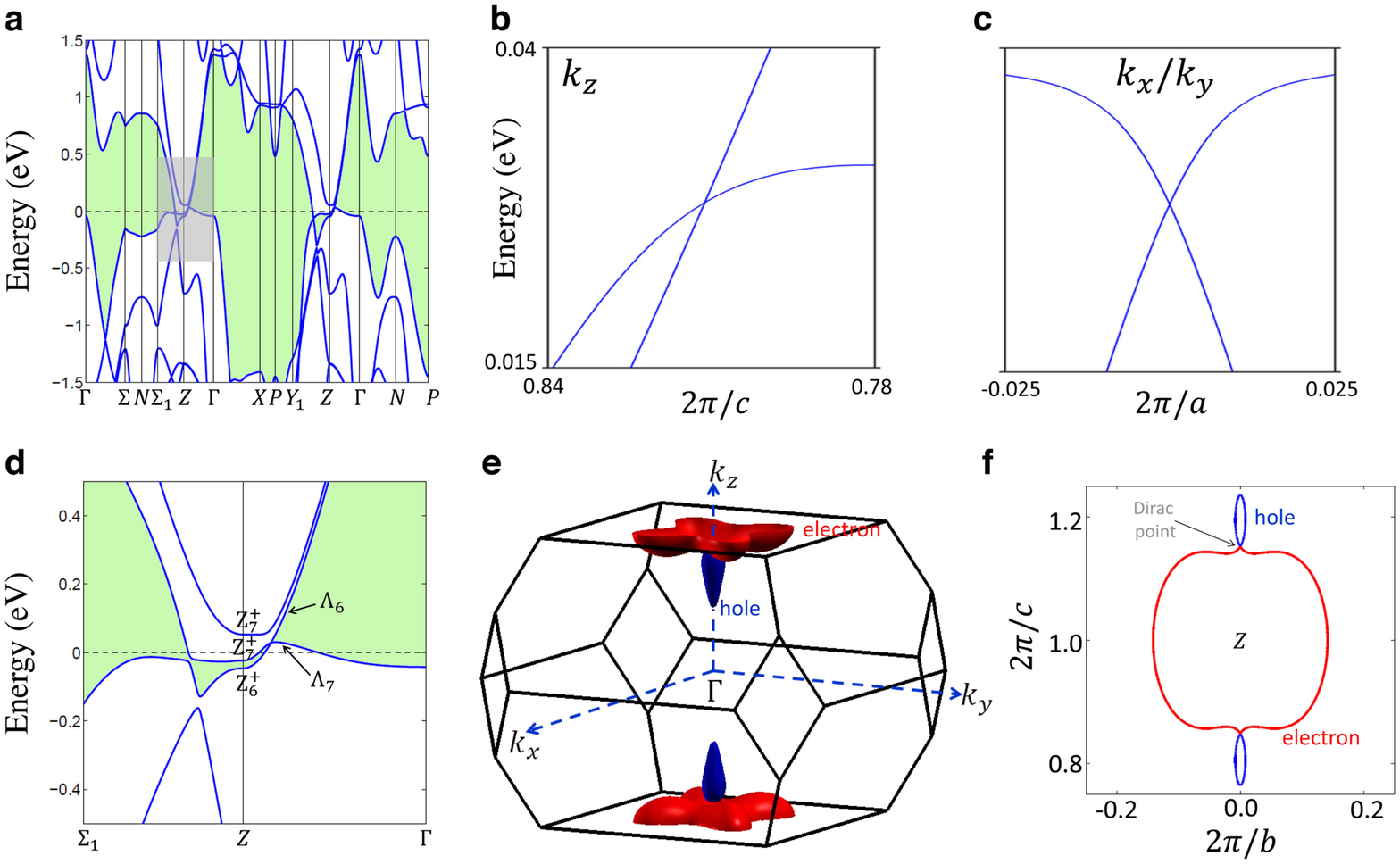}
\caption{
{\bf Type-II Dirac fermions in the bulk band structure of VAl$_3$.}
{\bf a,} The calculated bulk band structure of VAl$_3$ in the presence of spin-orbit coupling. The green shaded region shows the energy gap between the lowest conduction and valence bands. {\bf b, c} A zoomed-in calculation of the band dispersion along the $k_z$ (b) and $k_{x}$/$k_{y}$ (c) directions in the vicinity of the type-II Dirac node. {\bf d,} A zoomed-in view of the area highlighted by the gray box in (a). Each of the two crossing bands is marked with their corresponding space group representations $\Lambda_6$ and $\Lambda_7$. Because the two bands touching have different space group representations, they are prevented from hybridizing and gapping out. The type-II character of the Dirac node is clearly shown in this plot (i.e. there are both electron and hole bands touching at the energy of the Dirac node). {\bf e,} Bulk Fermi surface of VAl$_3$ with the red and blue pockets denoting the electron and hole bands, respectively. {\bf f,} Bulk constant energy contour in $(k_y,k_z)$ space at $k_x=0$ and at the energy of the type-II Weyl nodes. An electron pocket (red contour) enclosing the $Z$ point and two hole pockets (blue contours) touch to form a pair of type-II Dirac nodes.}
\label{band}
\end{figure}

\newpage

\begin{figure}
\centering
\includegraphics[width=16cm]{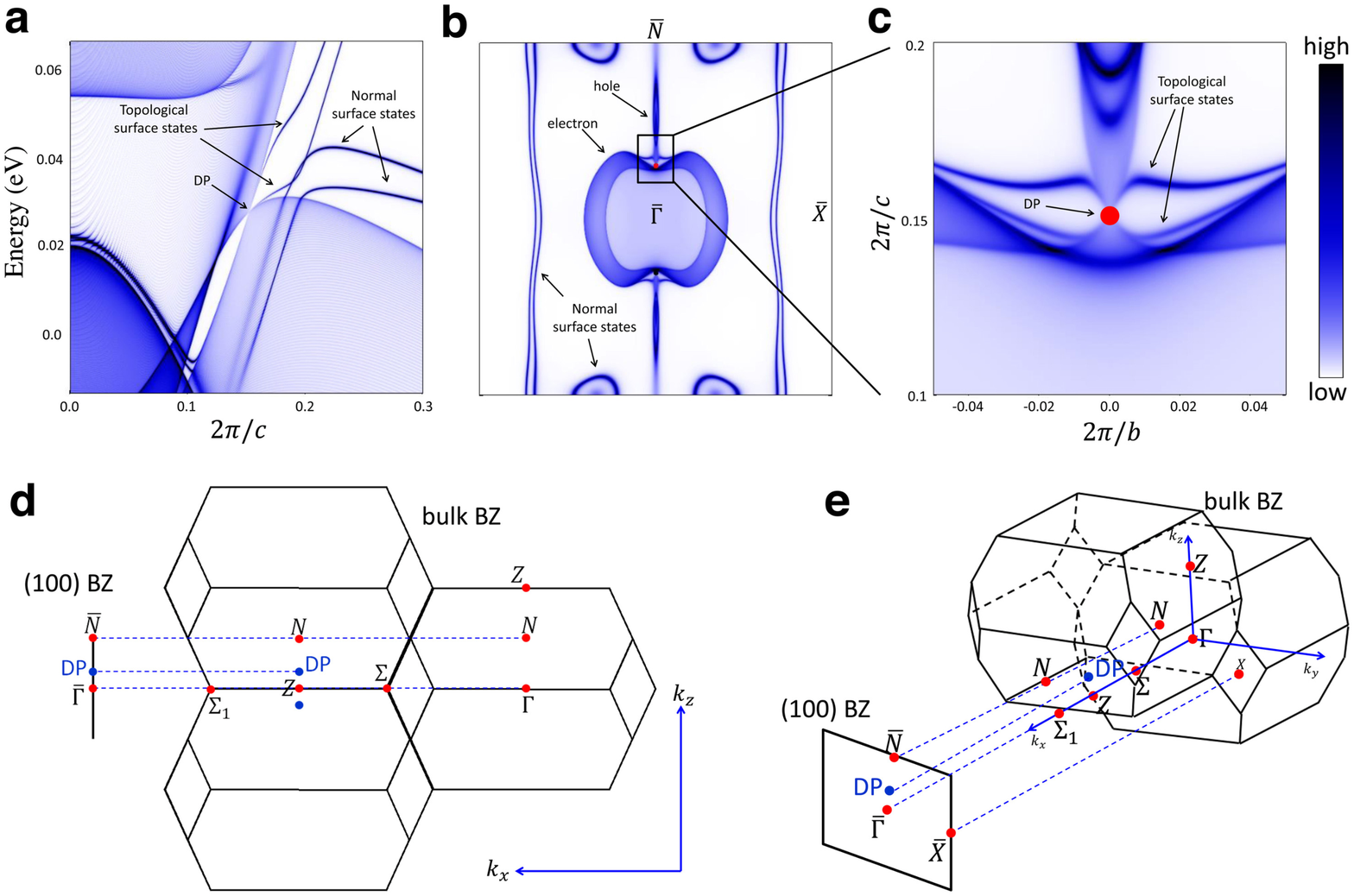}
\caption{
{\bf Fermi arc surface states associated with the type-II Dirac nodes.}
{\bf a,} Surface and bulk band structure of VAl$_3$ along $\bar{\Gamma}$-$\bar{N}$ direction on the (100) surface BZ. Two topological surface states are shown to be terminating directly onto the projected type-II Dirac node. The normal surface states are observed to avoid the Dirac node and merge into the bulk band continuum. {\bf b,} The calculated $k_{z}$-$k_{y}$ surface and bulk electronic structure at the energy of the bulk Dirac node ($\sim$ 28 meV above E$_F$). Enclosing $\bar{\Gamma}$ is a electron pocket and extending along $\bar{\Gamma}$-$\bar{N}$ is a hole pocket. The normal surface states are labeled with arrows and the projected type-II Dirac nodes are marked with red and black dots, which reside at the touching point between the electron and hole pockets along the $\bar{\Gamma}$-$\bar{N}$. {\bf c,} Zoomed-in view of the area highlighted by the black box that surrounds the projected type-II Dirac node in (b). Here we observe two pairs of topological surface states emerging from the Dirac node and merging into the bulk band continuum to connect the pair of Dirac nodes along $\bar{N}$-$\bar{\Gamma}$-$\bar{N}$. The fact that the topological surface states are pinned to the Dirac node is an interesting result because they are not constrained by the mirror Chern number to behave in this}
\label{ss}
\end{figure}
\addtocounter{figure}{-1}
\begin{figure*}[t!]
\caption{fashion and has a net zero Chiral charge. {\bf d,} The ($k_x$, $k_z$) side of three bulk Brillouin zones is shown to better understand the projection of Dirac nodes (blue dots) and relevant high-symmetry points (red dots) on the (100) surface BZ. {\bf e,} Similar to panel (d) but for two bulk Brillouin zones that are tilted in such a way that it becomes more clear where the projected Dirac nodes (blue dots) and relevant high-symmetry points (red dots) are positioned on the (100) surface BZ. As shown the Dirac nodes are projected along the $\bar{\Gamma}-\bar{N}$ high symmetry line.}
\label{ss}
\end{figure*}


\clearpage
\begin{figure}
\centering
\includegraphics[width=16cm]{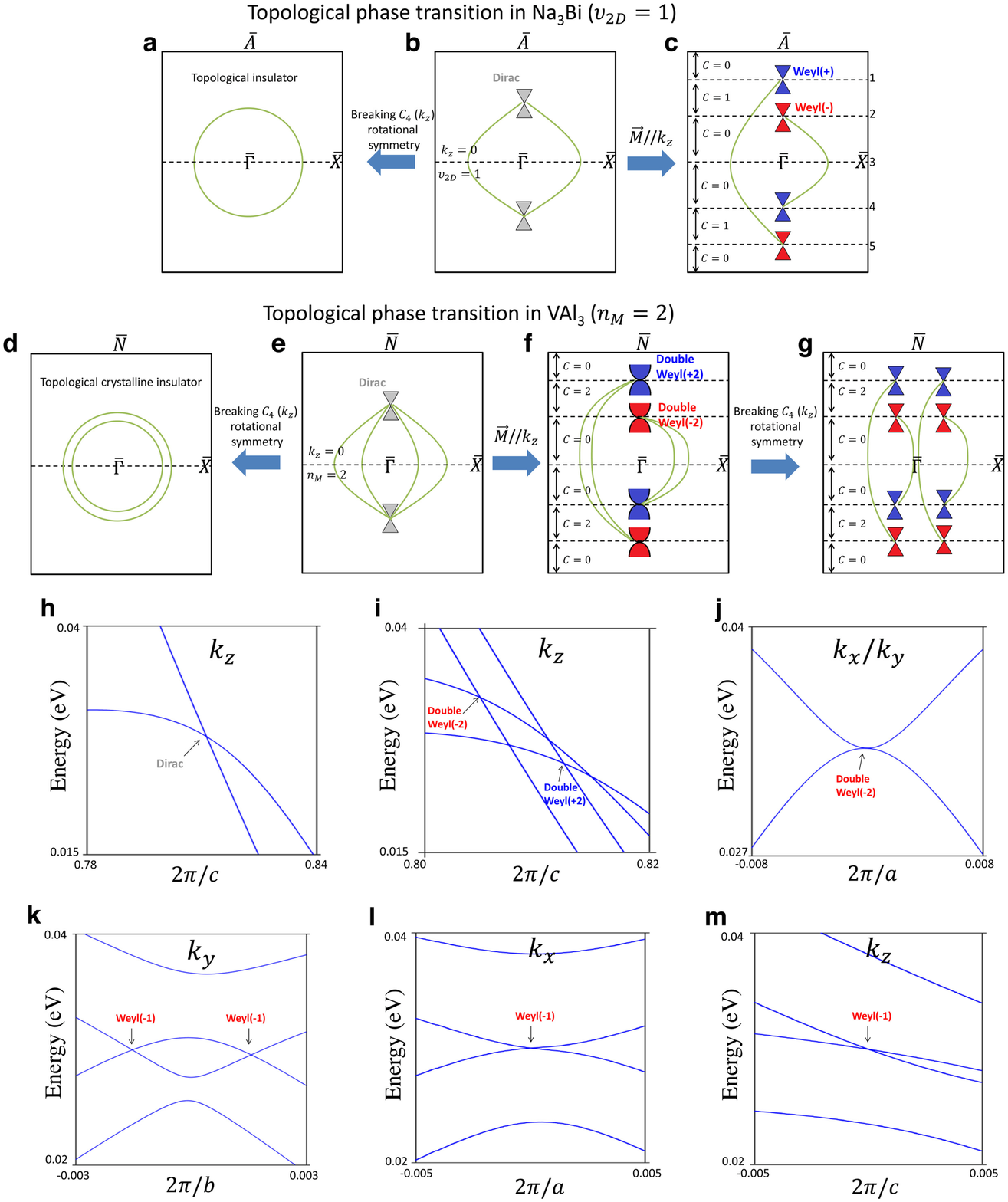}
\caption{{\bf Topological phase transitions in VAl$_3$.} {\bf a-c,} A schematic of the Fermi surface for Na$_3$Bi, (b), undergoing a transition to a topological insulating phase by breaking the $C_{3z}$ rotational symmetry, or a Weyl semimetal phase, (c), by applying an Zeeman field along the rotational symmetry preserving axis. {\bf d-e,} Topological phase transitions in VAl$_3$. \textbf{e}, an illustration of surface Fermi arcs and type-II Dirac nodes. The double Fermi arcs arise because of the defined $n_{M}=2$ mirror Chern number on the $k_{z}=0$-mirror plane.}
\label{FS}
\end{figure}

\addtocounter{figure}{-1}
\begin{figure*}[t!]
\caption{ \textbf{d,} Breaking the $C_{4z}$ rotational symmetry drives the type-II Dirac semimetal into a topological crystalline insulator phase, which results in two concentric topological surface states protected by the $\mathcal{M}_z$ symmetry, as shown in panel (d). \textbf{e,} Breaking time-reversal symmetry with an applied Zeeman field along the $k_{z}$-axis splits each Type-II Dirac node into a pair of double Weyl nodes, each with chiral charge $\pm{2}$. \textbf{f,} By further applying a $C_{4z}$ symmetry breaking perturbation, each double Weyl nodes split into two  singleWeyl nodes with $\pm{1}$ chiral charge. A net number of four Weyl nodes are generated from a singe type-II Dirac node. For panels (c), (f), and (g), the net Chern number for the regions defined by dashed lines is shown. {\bf h,} The calculated band dispersion of the Type-II Dirac cone in VAl$_3$ along the $k_z$ directions. {\bf i,} The calculated band dispersion in the presence of an Zeeman field. As shown, the type-II Dirac node splits into a pair (red and blue) type-II double Weyl cones with Weyl nodes of equal but opposite chiral charge $\pm{2}$. {\bf j,} The band dispersion along the $k_{x}$/$k_{y}$ direction for the type-II double Weyl cone with $-2$ chiral charge in panel (i). As expected, the Weyl node with $\pm{2}$ chiral charge is defined by the touching point between two quadratic bands. {\bf k-m,} By breaking the $C_{4z}$ symmetry, the double Weyl node in panel (j) splits into two single Weyl nodes with an equal chiral charge of $-1$. The zoomed-in band dispersions along $k_x$, $k_y$ and $k_z$ directions for the single Weyl nodes.}
\label{FS}
\end{figure*}

\clearpage
\begin{figure}
\centering
\includegraphics[width=12cm]{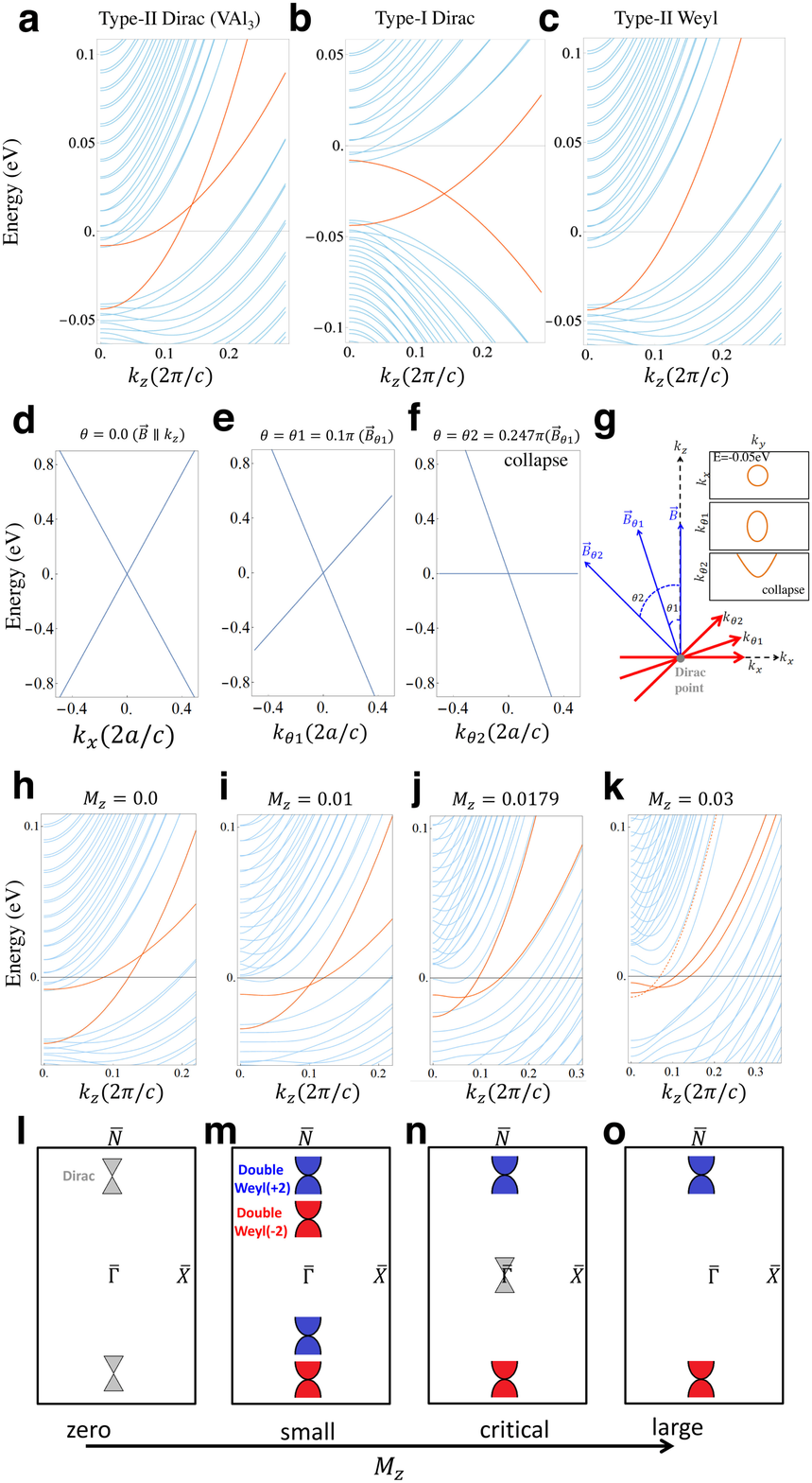}
\caption{{\bf Landau level spectrum of the type-II Dirac fermions in VAl$_3$.}}
\label{LL}
\end{figure}
\addtocounter{figure}{-1}
\begin{figure*}[t!]
\caption{ \textbf{a,} Landau level spectrum of the type-II Dirac fermions in VAl$_3$. The magnetic field is along the tilting direction of the type-II Dirac fermions, which is the $k_z$ direction for VAl$_3$. \textbf{b,c,} Same as panels (a,b) but for type-I Dirac fermions and type-II Weyl fermions. \textbf{d-f,} We vary the magnetic field direction within the $(k_x,k_z)$ plane. $\theta$ defines the angle between the magnetic field and $k_z$. The dispersions in panels (d-f) are along the $k$ space directions ($k_x, k_{\theta1}, k_{\theta2}$) shown in panel (g). They are within the $(k_x,k_z)$ plane and at the same time they are perpendicular to the magnetic field directions. \textbf{g,} Schematic illustration of different magnetic field directions and also the $k$ directions ($k_x, k_{\theta1}, k_{\theta2}$) that are perpendicular to the magnetic fields. The three insets on the top-right corner shows the constant energy contours in the $k$ plane that is perpendicular to the magnetic field. \textbf{h-k,} Landau level spectra of the type-II Dirac fermions in VAl$_3$ under different Zeeman couplings. The lowest Landau levels change from two counter-propagating chiral bands to two co-propagating chiral bands. This corresponds to the annihilation of the two quadratic double Weyl fermions at the $\bar{\Gamma}$ point as shown in panels (l-o).}
\label{LL}
\end{figure*}

\clearpage
\renewcommand{\thefigure}{S\arabic{figure}}
\setcounter{figure}{0}
\textbf{
\begin{center}
{\large \underline{Supplementary Information}:}
\end{center}
}

\vspace{0.2cm}

{\bf VAl3 family materials}

\begin{figure}[h]
\centering
\includegraphics[width=17cm]{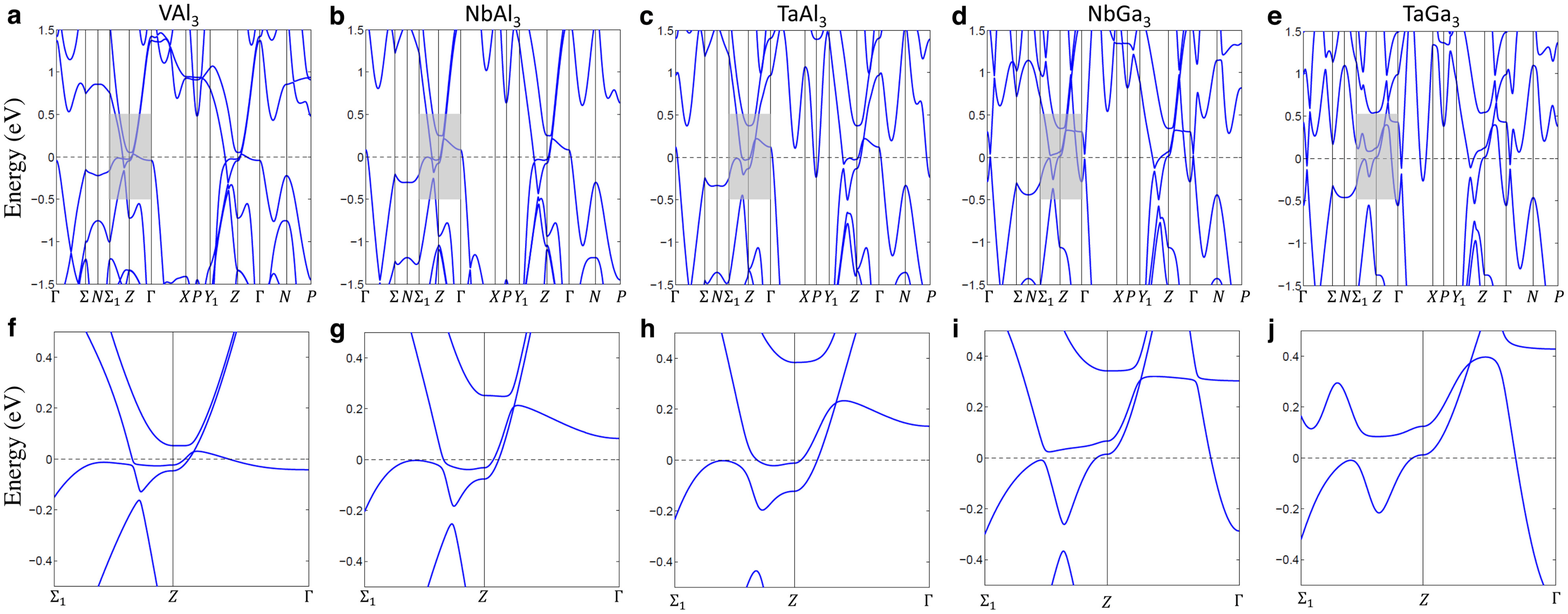}
\caption{
{\bf Band structure of VAl$_3$ family.}
{\bf a,} Calculated bulk band structure of VAl$_3$ with the inclusion of SOC.
{\bf b,} Calculated bulk band structure of NbAl$_3$ with the inclusion of SOC.
{\bf c,} Calculated bulk band structure of TaAl$_3$ with the inclusion of SOC.
{\bf d,} Calculated bulk band structure of NbGa$_3$ with the inclusion of SOC.
{\bf e,} Calculated bulk band structure of TaGa$_3$ with the inclusion of SOC.
{\bf f,} A zoomed-in view of {\bf a} for the area highlighted by the gray box.
{\bf g,} A zoomed-in view of {\bf b} for the area highlighted by the gray box.
{\bf h,} A zoomed-in view of {\bf c} for the area highlighted by the gray box.
{\bf i,} A zoomed-in view of {\bf d} for the area highlighted by the gray box.
{\bf j,} A zoomed-in view of {\bf e} for the area highlighted by the gray box.
}
\label{Sband}
\end{figure}

\begin{figure}
\centering
\includegraphics[width=16cm]{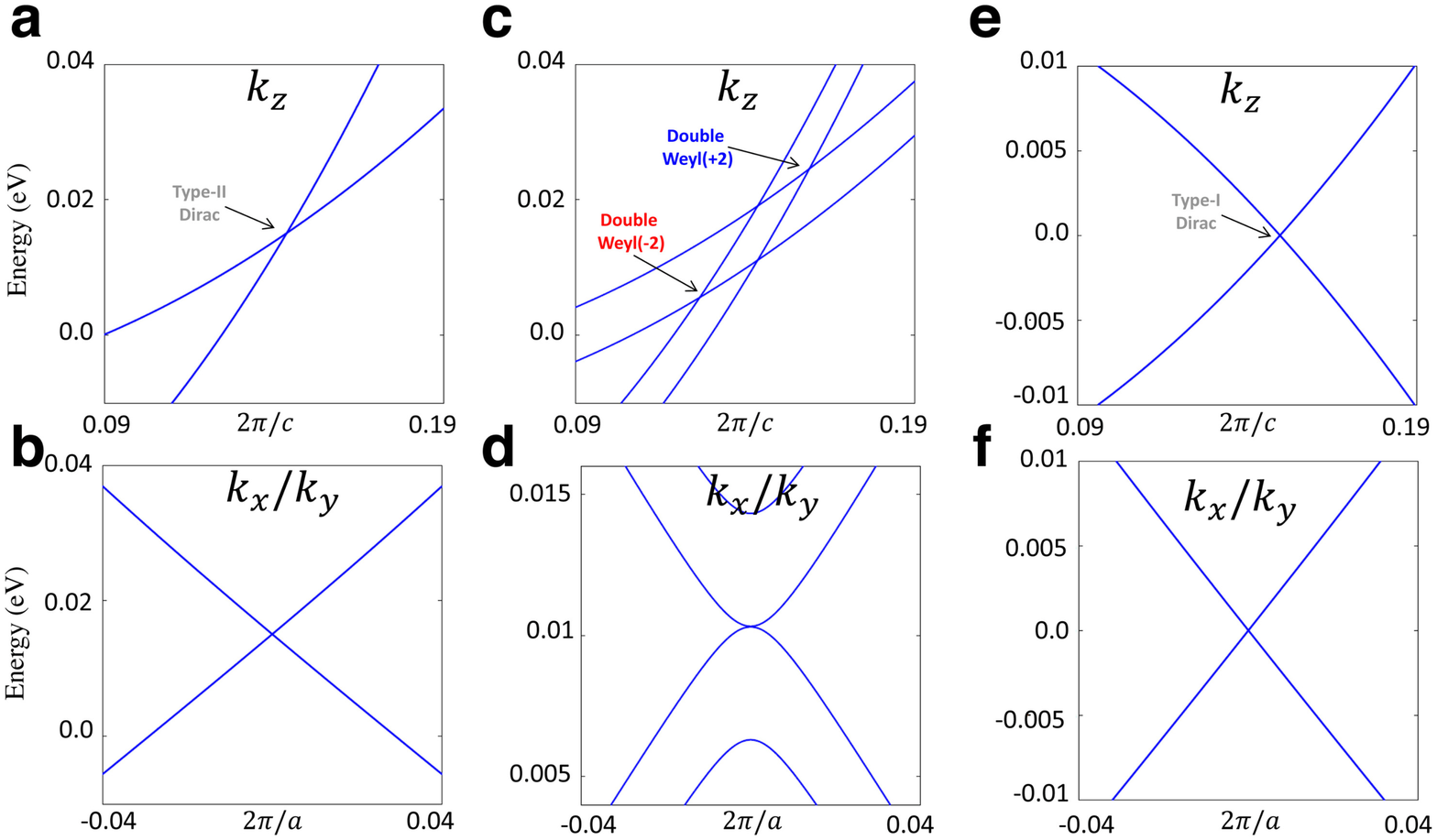}
\caption{
{\bf Band structure of} ${\bf k}\cdot{\bf p}$ {\bf model}
{\bf a,} Calculated band dispersion of $k \cdot p$ model along $k_z$ direction. The fitting parameters $\alpha_{0}$ = -0.0274, $\alpha_{5}$ = -0.0179, $\beta_{0}$ = 0.3730, $\beta_{5}$ = 0.1831, $\gamma_{0}$ = 2.0574, $\gamma_{5}$ = 0.8662, $\eta$ = 1.6590 - $i$0.7754, $\delta$ = -(0.8351 + $i$0.0301), and $\xi$ = -(0.0048 + $i$0.1961) were used in this work.
{\bf b,} Calculated band dispersion of $k \cdot p$ model along $k_x$ and $k_y$ directions.
{\bf c,} Calculated band dispersion of $k \cdot p$ model along $k_z$ direction under $k_z$ exchange field.
{\bf d,} Calculated band dispersion of $k \cdot p$ model along $k_x$ and $k_y$ directions under $k_z$ exchange field.
{\bf e,} Calculated band dispersion of $k \cdot p$ model along $k_z$ direction when $\alpha_{0}$ = $\gamma_{0}$ = 0.
{\bf f,} Calculated band dispersion of $k \cdot p$ model along $k_x$ and $k_y$ directions when $\alpha_{0}$ = $\gamma_{0}$ = 0.
}
\label{model}
\end{figure}

\clearpage
${\bf k}\cdot{\bf p}$ {\bf model.}

To better understanding the fundamental properties of Type-II Dirac semimetal, we synthesize a k.p model is as follow. Let us consider 4 $\times$ 4 matrix minimal Hamiltonian, which general has the following form,
\begin{equation}
 H({\bf k}) =
 \sum_{i,j=0}^{3}a_{ij}({\bf k})\sigma_{i}\tau_{j} = \\
 \left(
 \begin{array}{cc}
   h_{\uparrow\uparrow}({\bf k}) & h_{\uparrow\downarrow}({\bf k})  \\
   h_{\downarrow\uparrow}({\bf k}) & h_{\downarrow\downarrow}({\bf k})
 \end{array}
 \right)
\end{equation}
where the $\sigma_{0}$ and $\tau_{0}$ are identity matrices. $\sigma_{1,2,3}$ and $\tau_{1,2,3}$ are Pauli matrix which indicates spin and orbital degree of freedom, respectively. We assume $a_{ij}$({\bf k}) and $h_{\sigma \sigma^{\prime}}$ are real functions. To fit the VAl$_3$, we constrain the form of $H({\bf k})$ by time-reversal symmetry (TRS), spatial inversion symmetry (IS), and $C_4$ rotational symmetry along the z-axis. At first, we impose TRS on $H({\bf k})$. The TRS can be represented by the operator $\Theta$ = $i\sigma_{2}\kappa$, where $\kappa$ denotes complex conjugation and $\sigma_2$ is the second Pauli matrix acting on the electron spin. Under TRS, $H({\bf k})$ requires $\Theta H({\bf k})\Theta^{-1}$ = $H(-{\bf k})$, giving rise to $h_{\downarrow\downarrow}({\bf k})$ = $h^{*}_{\uparrow\uparrow}(-{\bf k})$ and $h_{\downarrow\uparrow}({\bf k})$ = -$h^{*}_{\uparrow\downarrow}(-{\bf k})$. Thus Hamiltonian can be written in the following form.
\begin{equation}
 H({\bf k}) =
 \left(
  \begin{array}{cc}
    h_{\uparrow\uparrow}({\bf k}) & h_{\uparrow\downarrow}({\bf k})  \\
   -h^{*}_{\downarrow\uparrow}(-{\bf k}) & h^{*}_{\downarrow\downarrow}(-{\bf k})
  \end{array}
 \right)
\end{equation}
Second, we impose IS on $H({\bf k})$. The invariance of the $H({\bf k})$ under IS, that is, $PH({\bf k})P^{-1}$, where $P$ is IS operator. Since $P$ is independent of the spin rotation, $P$ can be written $P$ = $\alpha_{0}\tau_{0}$ + \bm{$\alpha$}$\cdot$\bm{$\tau$}, where $\alpha_{1,2,3,4}$ are complex numbers. In addition, we claim because $P^{2}$ = 1 because it should be equivalent to the identity operator up to a global U(1) phase factor, that is, $P^2$ = $e^{i2\phi} \cdot {\bf I}$ = $\alpha^{2}_{0}$ + $\bm{\alpha}^2$+2$\alpha_{0}$ \bm{$\alpha$}$\cdot$\bm{$\tau$}. Therefore, $P$ = $\pm e^{i\phi}\tau_{0}$ or $e^{i\pi}$\bm{$\alpha^{\prime}$}$\cdot$\bm{$\tau$}, where \bm{$\alpha^{\prime}$}$^2$ = 1. To determine \bm{$\alpha^{\prime}$} and $\phi$, three conditions constrain are needed, (1) $[\Theta, P]$ = 0, (2) $P^{\dagger}P$ = 1, and (3) $(\Theta P)^2$ = -1. Thus, the general solution of $P$ is $P$ = $\pm\tau_{0}$ or $cos\theta\tau_{3}$ –- $sin\theta\tau_{1}$ with $\theta\in[0, 2\pi]$. In this work, we pick
up $P$ = $\tau_{0}$. Finally, let us consider rotational symmetry on $H$({\bf k}). We choose $k_z$ as a rotation axis and $C_4$ matrix are diagonal $C_4$ = $diag(u_{A\uparrow}, u_{B\uparrow}, u_{A\downarrow}, u_{B\downarrow})$ = $diag(\alpha_p, \alpha_q, \alpha_r, \alpha_s)$, where $\alpha_p$ = $e^{i\frac{\pi}{2}(p+\frac{1}{2})}$ with $p$ = 0, 1, 2, 3. Thus $C_4$ can be written in following form.
\begin{equation}
 C_{4} =
 \left(
  \begin{array}{cc}
    e^{i\pi(\frac{1+p+q}{4})+\frac{p-q}{4}\tau_{3}} & 0  \\
    0 & e^{i\pi(\frac{1+r+s}{4})+\frac{r-s}{4}\tau_{3}}
  \end{array}
 \right)
\end{equation}
The $H({\bf k})$ under $C_4$ symmetry results in $C_4 H(k_+, k_-, k_z) C^{-1}_4$ = $H(k_{+}e^{i\frac{\pi}{2}}, k_{-}e^{-i\frac{\pi}{2}}, kz)$. where $k_{\pm}$ = $k_x$ $\pm ik_y$. From this invariance, we can obtain
\begin{eqnarray}
    & e^{i\frac{\pi}{4}(p-q)\tau_{3}}h_{\uparrow\uparrow}(k_+, k_-, k_z)e^{-i\frac{\pi}{4}(p-q)\tau_{3}} = h_{\uparrow\uparrow}(k_+e^{i\frac{\pi}{2}}, k_-e^{-i\frac{\pi}{2}}, k_z),
    \nonumber\\
    & e^{i\frac{\pi}{4}(p+q-r-s)}e^{i\frac{\pi}{4}(p-q)\tau_{3}}h_{\uparrow\downarrow}(k_+, k_-, k_z)e^{-i\frac{\pi}{4}(r-s)\tau_{3}} = h_{\uparrow\downarrow}(k_+e^{i\frac{\pi}{2}}, k_-e^{-i\frac{\pi}{2}}, k_z)
\end{eqnarray}
Since $[\Theta, C_4]$ =0, $C_4$ can be written $C_4$ = $diag(\alpha_p, \alpha_q, \alpha^*_p, \alpha^*_q)$.
Therefore eq.(4) becomes
\begin{eqnarray}
 & e^{i\frac{\pi}{4}(p-q)\tau_{3}}h_{\uparrow\uparrow}(k_+, k_-, k_z)e^{-i\frac{\pi}{4}(p-q)\tau_{3}} = h_{\uparrow\uparrow}(k_+e^{i\frac{\pi}{2}}, k_-e^{-i\frac{\pi}{2}}, k_z),
    \nonumber\\
 & e^{i\frac{\pi}{4}(p-r)}e^{i\frac{\pi}{4}(p-q)\tau_{3}}h_{\uparrow\downarrow}(k_+, k_-, k_z)e^{-i\frac{\pi}{4}(p-q)\tau_{3}} = h_{\uparrow\downarrow}(k_+e^{i\frac{\pi}{2}}, k_-e^{-i\frac{\pi}{2}}, k_z)
\end{eqnarray}
In terms of \bm{$\tau$} and $h_{\sigma \sigma^{\prime}}$ can be expressed as
\begin{eqnarray}
 & h_{\uparrow\uparrow}({\bf k}) = f_{0}({\bf k}) + f_{+}({\bf k})\tau_{+} + f^*_{+}({\bf k})\tau_{-} + f_{z}({\bf k})\tau_{3},
 \nonumber\\
 & h_{\uparrow\downarrow}({\bf k}) = g_{0}({\bf k}) + g_{+}({\bf k})\tau_{+} + g^*_{-}({\bf k})\tau_{-} + g_{z}({\bf k})\tau_{3}
\end{eqnarray}
where $\tau_{\pm}$ = $\tau_{1}\pm i\tau_{2}$, and due to hermitian requirement, $f_{0,z}$ are real functions and $f_{+}$, $g_{0,\pm,z}$ are complex functions.
From eq.(5) and eq.(6), we obtain
\begin{eqnarray}
 & f_{0}(k_{+}, k_{-}, k_{z}) = f_{0}(k_{+}e^{i\frac{\pi}{2}}, k_{-}^{-i\frac{\pi}{2}}, k_{z}),
 \nonumber\\
 & f_{z}(k_{+}, k_{-}, k_{z}) = f_{z}(k_{+}e^{i\frac{\pi}{2}}, k_{-}^{-i\frac{\pi}{2}}, k_{z}),
 \nonumber\\
 & e^{i\frac{\pi}{2}(p-q)}f_{+}(k_{+}, k_{-}, k_{z}) = f_{+}(k_{+}e^{i\frac{\pi}{2}}, k_{-}^{-i\frac{\pi}{2}}, k_{z})
\end{eqnarray}
and
\begin{eqnarray}
 & e^{i\frac{\pi}{2}(p-r)}(g_{0}+g_{z})(k_{+}, k_{-}, k_{z}) = (g_{0}+g_{z})(k_{+}e^{i\frac{\pi}{2}}, k_{-}^{-i\frac{\pi}{2}}, k_{z}),
 \nonumber\\
 & e^{i\frac{\pi}{2}(q-s)}(g_{0}-g_{z})(k_{+}, k_{-}, k_{z}) = (g_{0}-g_{z})(k_{+}e^{i\frac{\pi}{2}}, k_{-}^{-i\frac{\pi}{2}}, k_{z}),
 \nonumber\\
 & e^{i\frac{\pi}{2}(q-r)}g_{\pm}(k_{+}, k_{-}, k_{z}) = g_{\pm}(k_{+}e^{i\frac{\pi}{2}}, k_{-}^{-i\frac{\pi}{2}}, k_{z}),
\end{eqnarray}

Combining with TRS and IS, $H({\bf k}) = \sum_{i,j=0}^{3}a_{ij}({\bf k})\sigma_{i}\tau_{j}$ becomes $H({\bf k}) = \sum_{i=0}^{5}\widetilde{a}_{i}({\bf k})\Gamma_{i}$, where $\Gamma_0$ = $\sigma_{0}\tau_{0}$, $\Gamma_1$ = $\tau_{1}$, $\Gamma_2$ = $\sigma_{3}\tau_{2}$, $\Gamma_3$ = $\sigma_{1}\tau_{2}$, $\Gamma_4$ = $\sigma_{2}\tau_{2}$, and $\Gamma_5$ = $\tau_{3}$, and $\widetilde{a}_{i}$ are real and even functions of ${\bf k}$. Hence $h_{\uparrow\uparrow}$ = $\widetilde{a}_{0}$ + $\widetilde{a}_{1}\tau_{1}$ + $\widetilde{a}_{2}\tau_{2}$ + $\widetilde{a}_{5}\tau_{3}$ and $h_{\uparrow\downarrow}$ = ($\widetilde{a}_{3}$ - $i\widetilde{a}_{4}$)$\tau_2$. Imposing $C_4$ constrain, the form of $H({\bf k})$ becomes $H(0, 0, k_z) \sim \sum_{i=0}\widetilde{a}_{i}(0, 0, k_z)\Gamma_{i}$ with $\Gamma_{i}$ along $k_z$ axis only include $\Gamma_{0}$ and $\Gamma_{5}$. When $P=\tau_0$, we can  obtain $\widetilde{a}_{i}(0, 0, k_z)$ = $\widetilde{a}_{i}(0, 0, -k_z)$ for $i$=0, 5 and  $\widetilde{a}_{i}(0, 0, k_z)$ =0 for $i$
= 1, 2, 3,4. Thus $h_{\uparrow\uparrow}$ and $h_{\uparrow\downarrow}$ can be written in terms of $f$ and $g$ functions.
\begin{eqnarray}
 & h_{\uparrow\uparrow} = f_{0}({\bf k}) + f_{+}({\bf k})\tau_{+} + f^*_{+}({\bf k})\tau_{-} + f_{z}({\bf k})\tau_{3},
 \nonumber\\
 & h_{\uparrow\downarrow} = g_{0}({\bf k}) + g_{+}({\bf k})\tau_{+} + g_{-}({\bf k})\tau_{-} + g_{z}({\bf k})\tau_{3}
\end{eqnarray}
where $f_{0}({\bf k})$ = $\widetilde{a}_{0}({\bf k})$, $f_{+}({\bf k})$ = ($\widetilde{a}_{1}({\bf k})$ - $i\widetilde{a}_{2}({\bf k})$)/2, $f^*_{+}({\bf k})$ = ($\widetilde{a}_{1}({\bf k})$ + $i\widetilde{a}_{2}({\bf k})$)/2, $f_{z}({\bf k})$ = $\widetilde{a}_{5}({\bf k})$, $g_{+}({\bf k})$ = $-i$($\widetilde{a}_{3}({\bf k})$ - $i\widetilde{a}_{4}({\bf k})$)/2, $g_{-}({\bf k})$ = $i$($\widetilde{a}_{3}({\bf k})$ - $i\widetilde{a}_{4}({\bf k})$)/2, and  $g_{0}({\bf k})$ = $g_{z}({\bf k})$ = 0.
According to eq.(7) and eq.(8), we have
\begin{eqnarray}
 &\widetilde{a}_{0}(k_+, k_-, k_z) = \widetilde{a}_{0}(k_{+}e^{i\frac{\pi}{2}}, k_{-}e^{i\frac{\pi}{2}}, k_z),
 \nonumber\\
 &\widetilde{a}_{5}(k_+, k_-, k_z) = \widetilde{a}_{5}(k_{+}e^{i\frac{\pi}{2}}, k_{-}e^{i\frac{\pi}{2}}, k_z),
 \nonumber\\
 &e^{i\frac{\pi}{2}(p-q)}f_{+} = \frac{1}{2}[\widetilde{a}_{1}(k_{+}e^{i\frac{\pi}{2}}, k_{-}e^{i\frac{\pi}{2}}, k_z)-i\widetilde{a}_{2}(k_{+}e^{i\frac{\pi}{2}}, k_{-}e^{i\frac{\pi}{2}}, k_z)],
 \nonumber\\
 &e^{i\frac{\pi}{2}(q-r)}g_{\pm} = \mp\frac{i}{2}[\widetilde{a}_{3}(k_{+}e^{i\frac{\pi}{2}}, k_{-}e^{i\frac{\pi}{2}}, k_z)-i\widetilde{a}_{4}(k_{+}e^{i\frac{\pi}{2}}, k_{-}e^{i\frac{\pi}{2}}, k_z)]
\end{eqnarray}
As $f_{+}(0, 0, k_{z})$ = $g_{\pm}(0, 0, k_{z})$ = 0, $e^{i\frac{\pi}{2}(p-q)}$ and $e^{i\frac{\pi}{2}(q-r)}$ cannot be 1. In other words, $\alpha_{p} \ne \alpha_{q}$ and $\alpha_{r} \ne \alpha_{q}$. Hence $\alpha_{p}$, $\alpha_{q}$, $\alpha^*_{p}$, and $\alpha^*_{q}$ are different rotational eigenvalues.
Now the $\widetilde{a}_{i}$, $f$ and $g$ functions can be determined.
\begin{eqnarray}
 \widetilde{a}_{0,5} = \alpha + \beta k_{+}k_{-} + \gamma k^2_{z},
 \nonumber\\
 f_{+}(k_+, k_-, k_z) = \eta k_{+}k_{z},
 \nonumber\\
 g_{+}(k_+, k_-, k_z) = \delta k^2_{+} + \xi k^2_{-},
 \nonumber\\
 g_{-}(k_+, k_-, k_z) = -g_{+}(k_+, k_-, k_z)
\end{eqnarray}
where $\alpha$, $\beta$, $\gamma$ and $\eta$, $\delta$, $\xi$ are real numbers and complex numbers, respectively.
Therefore $H({\bf k})$ can be written in the following form.
\begin{equation}
 H({\bf k}) =
 \left(
 \begin{array}{cccc}
   D_{1} & 2\eta k_{+}k_{z} & 0 & 2(\delta k^2_{+} + \xi k^2_{-}) \\
   2\eta^* k_{-}k_{z} & D_2 & -2(\delta k^2_{+} + \xi k^2_{-}) & 0 \\
   0 & -2(\delta^* k^2_{-} + \xi^* k^2_{+}) & D_1 & 2\eta^* k_{-}k_{z} \\
   2(\delta^* k^2_{-} + \xi^* k^2_{+}) & 0 & 2\eta k_{+}k_{z} & D_2
 \end{array}
 \right)
\end{equation}
where $D_{1}$ = $(\alpha_{0} + \alpha_{5}) + (\beta_{0} + \beta_{5})k_{+}k_{-} + (\gamma_{0} + \gamma_{5})k^2_{z}$, and $D_{2}$ = $(\alpha_{0} - \alpha_{5}) + (\beta_{0} - \beta_{5})k_{+}k_{-} + (\gamma_{0} - \gamma_{5})k^2_{z}$

Figure S\ref{model} shows the band structure of $k \cdot p$ model. The fitting parameters $\alpha_{0}$ = -0.0274, $\alpha_{5}$ = -0.0179, $\beta_{0}$ = 0.3730, $\beta_{5}$ = 0.1831, $\gamma_{0}$ = 2.0574, $\gamma_{5}$ = 0.8662, $\eta$ = 1.6590 - $i$0.7754, $\delta$ = -(0.8351 + $i$0.0301), and $\xi$ = -(0.0048 + $i$0.1961) were used to diagonalize $H({\bf k})$. Our calculation shows a clear Type-II Dirac band dispersion with a four-fold degenerate Dirac point (Fig. S\ref{model}(a)). We also check the band structures along $k_x$ and $k_y$ directions in the vicinity of gapless points and found that the bands disperse linearly along all directions (Fig. S\ref{model}(b)), which consistent very well with our DFT results. In the presence of exchange field $M_z\sigma_{3}$, two doubly degenerate Dirac bands were split into four singly degenerate bands. The band dispersion along $k_z$ direction shows Type-II Weyl band dispersion (Fig. S\ref{model}(c)) and depicts double Weyl quadratic band dispersion along $k_x$ and
$k_y$ directions (Fig. S\ref{model}(d)). Furthermore, our model not only catch the main features of VAl$_3$ but also providing a platform to study Type-I Dirac and Weyl band dispersion. Setting $\alpha_{0}$ = $\gamma_{0}$ = 0, our model presents a Type-I and dispersion along all three directions (Fig. S\ref{model}(e) and (f)).


\end{document}